# Fetal cardiovascular decompensation during labor predicted from the individual heart rate: a prospective study in fetal sheep near term and the impact of low sampling rate

Nathan Gold[1], Christophe L. Herry[2], Xiaogang Wang[1,3], Martin G. Frasch[4]

[1] Department of Mathematics and Statistics, York University, Toronto, ON, Canada
[2] Dynamical Analysis Laboratory, Clinical Epidemiology Program, Ottawa Hospital Research Institute, Ottawa, ON, Canada
[3] Institute of Big Data, Qing Hua University, Beijing, China
[4] Department of Obstetrics and Gynecology and Center on Human Development and Disability, University of Washington, Seattle, WA, USA

**Address of correspondence:**

Martin G. Frasch
Department of Obstetrics and Gynecology
University of Washington
1959 NE Pacific St
Box 356460
Seattle, WA 98195
Phone: +1-206-543-5892
Fax: +1-206-543-3915
Email: mfrasch@uw.edu

**Precis**
- We present a novel computerized fetal heart rate (FHR) intrapartum algorithm for early and individualized prediction of fetal cardiovascular decompensation ("the event"), a key event in the causal chain leading to brain injury
- This real time machine learning algorithm performs well on noisy FHR data and requires ~2 hours to train on the individual FHR tracings in the first stage of labor; once trained, the algorithm predicts the event with 92% sensitivity
- We show that the algorithm's performance suffers reducing sensitivity to 67% when the FHR is acquired at the sampling rate of 4 Hz used in the ultrasound (CTG) monitors compared to the ECG-derived signal as it can be acquired from maternal abdominal ECG

**Abstract**
**Objective.** When exposed to repetitive umbilical cord occlusions (UCO) with worsening acidemia, fetuses eventually develop cardiovascular decompensation manifesting as pathological arterial blood pressure (ABP) responses to fetal heart rate (FHR) decelerations. Failure to maintain cardiac output during labor is a key event leading up to brain injury. We reported that the timing when fetuses demonstrate this cardiovascular phenotype is highly individual and was impossible to predict. We hypothesized that this phenotype would be reflected in the individual behavior of heart rate variability (HRV) as measured by root mean square of successive differences of R-R intervals (RMSSD), a measure of vagal modulation of HRV, which is known to increase with worsening acidemia. This is clinically relevant, because HRV can computed in real time intrapartum. Consequently, we expected to predict the individual timing of the event when hypotensive ABP pattern would emerge in a fetus from a series of continuous RMSSD data.
**Methods.** Fourteen near-term fetal sheep were chronically instrumented with vascular catheters to record fetal arterial blood pressure, umbilical cord occluder to mimic uterine contractions occurring during human labour and ECG electrodes to record the ECG-derived HRV measure RMSSD. All animals were studied over a ~6 hour period. After a 1-2 hour baseline control period, the animals underwent mild, moderate, and severe series of repetitive UCO. We applied the recently developed change point algorithm to detect physiologically meaningful changes in RMSSD dynamics with worsening acidemia. To mimic clinical scenarios using ultrasound-based 4 Hz FHR sampling rate, we recomputed RMSSD from 4 Hz sampled ECG and compared the performance of our algorithm under both conditions (1000 Hz versus 4 Hz).
**Results.** The RMSSD values were highly non-stationary, with four different regimes and three regime changes, corresponding to a baseline period followed by mild, moderate and severe UCO series. Each time series was characterized by seemingly randomly occurring (in terms of timing of the individual onset) increase in RMSSD values at different time points during the moderate UCO series and at the start of the severe UCO series. This event manifested as an increasing trend in RMSSD values, which counter-intuitively emerged as a period of relative stationarity for the time series. Our algorithm identified these change points as the individual time points of cardiovascular decompensation with 92% sensitivity, 86% accuracy and 92% precision which corresponded to 14 ± 21 minutes before the visual identification. In 4 Hz RMSSD time series, the algorithm detected the event with 3 times earlier detection times than at 1000 Hz, i.e., producing false positive alarms with 67% sensitivity, 14% accuracy and 18% precision. We identified the overestimation of baseline FHR variability by RMSSD at 4 Hz sampling rate to be the cause of this phenomenon.
**Conclusions.** Our findings validated the hypothesis that our HRV-based algorithm identifies individual time points of ABP responses to UCO with worsening acidemia by extracting change point information from the physiologically related fluctuations in the RMSSD time series. This performance depends on the acquisition accuracy of beat to beat fluctuations achieved in trans-abdominal ECG devices and fails at the sampling rate used clinically in ultrasound-based systems.

**Objective**

Electronic fetal monitoring (EFM) fails to identify fetuses at risk of incipient brain injury. The efforts to identify intrapartum acidemia using EFM have failed, in particular using fetal heart rate (FHR) monitoring, because fetal brain injury is poorly correlated with acidemia.[1] Brain compromise due to hypoxia-ischemia (HI) can ensue when the fetal cerebral blood flow is persistently reduced, e.g. due to precipitous drop in cerebral perfusion pressure resulting from cardiovascular decompensation.[2,3] Bezold-Jarisch reflex (BJR) is a vagal depressor reflex observed in fetal sheep under the conditions of umbilical cord occlusions (UCO) with worsening acidemia which leads to cardiovascular decompensation.[4] We asked whether FHR monitoring can capture the BJR-mediated vagal sensing of acidemia. We studied the relationship between fetal systemic arterial blood pressure (ABP) and FHR in an animal model of human labor. We had reported that fetuses have individual cardiovascular phenotype in their responses to increasing acidemia due to repetitive intermittent hypoxia.[3] We hypothesized that such phenotype would be reflected in individual responses of heart rate variability (HRV) as measured by root mean square of successive differences of R-R intervals (RMSSD), a measure of vagal modulation of HRV known to increase with worsening acidemia.[5–7] Consequently, a series of continuously computed RMSSD data will consistently predict the event when a hypotensive ABP pattern emerges in an individual fetus.[3] Because the current standard of EFM depends on ultrasound-based FHR monitoring, we also tested the impact of its inherently lower FHR sampling rate precision of 4 Hz versus the golden standard electrocardiogram (ECG) - derived 1000 Hz on the ability to individually predict cardiovascular decompensation.

**Study design**

Experimental methods and data acquisition have been presented elsewhere.[8] Briefly, fourteen near-term fetal sheep were chronically instrumented with vascular catheters to record fetal arterial blood pressure, umbilical cord occluder to mimic uterine contractions occurring during human labour and ECG electrodes to record ECG-derived HRV measure RMSSD. Animal care followed the guidelines of the Canadian Council on Animal Care and was approved by the University of Western Ontario Council on Animal Care.

Surgical preparation

Fourteen near-term ovine fetuses (123 ± 2 days gestational age (GA), term = 145 days) of mixed breed were surgically instrumented. The anesthetic and surgical procedures and postoperative care of the animals have been previously described.[3,9] Briefly, polyvinyl catheters were placed in the right and left brachiocephalic arteries, the cephalic vein, and the amniotic cavity. Stainless steel electrodes were sewn onto the fetal chest to monitor the electrocardiogram (ECG). A polyvinyl catheter was also placed in the maternal femoral vein. Stainless steel electrodes were additionally implanted biparietally on the dura for the recording of electrocorticogram, ECOG, as a measure of summated brain electrical activity (results reported elsewhere [3,8,10]). An inflatable silicon rubber cuff (In Vivo Metric, Healdsburg, CA) for UCO induction was placed around the proximal portion of the umbilical cord and secured to the abdominal skin. Once the fetus was returned to the uterus, a catheter was placed in the amniotic fluid cavity. Antibiotics were administered intravenously to the mother (0.2 g of trimethoprim and 1.2 g sulfadoxine, Schering Canada Inc., Pointe-Claire, Canada) and fetus and into the amniotic cavity (1 million IU penicillin G sodium, Pharmaceutical Partners of Canada, Richmond Hill, Canada). Amniotic fluid lost during surgery was replaced with warm saline. The uterus and abdominal wall incisions were sutured in layers and the catheters exteriorized through the maternal flank and secured to the back of the ewe in a plastic pouch.

Postoperatively, animals were allowed four days to recover prior to experimentation and daily antibiotic administration was continued intravenously to the mother (0.2 g trimethoprim and 1.2 g sulfadoxine), into the fetal vein and the amniotic cavity (1 million IU penicillin G sodium, respectively). Arterial blood was sampled for evaluation of the maternal and fetal condition and catheters were flushed with heparinized saline to maintain patency. Animals were 130 ± 1 day GA on the first day of the experimental study.

Experimental procedure

All animals were studied over a ~6 hour period. Fetal chronic hypoxia was defined as arterial $O_2$Sat <55% as measured on postoperative days 1 to 3 and at baseline prior to beginning the UCOs. The first group comprised five fetuses that were also spontaneously hypoxic (n=5, H/UCO). The second group of fetuses was normoxic ($O_2$Sat>55% before UCOs) (n=9, N/UCO). The experimental protocol has been reported.[7,9,11] After a 1-2 hour baseline control period, the animals underwent mild, moderate, and severe series of repetitive UCOs by graduated inflation of the occluder cuff with a saline solution. During the first hour following the baseline period, mild variable FHR decelerations were performed with a partial UCO for 1 minute duration every 2.5 minutes, with the goal of decreasing FHR by ~ 30 bpm, corresponding to a ~ 50% reduction in umbilical blood flow.[12,13] During the second hour, moderate variable FHR decelerations were performed with increased partial UCO for 1 minute duration every 2.5 minutes with the goal of decreasing FHR by ~ 60 bpm, corresponding to a ~ 75% reduction in umbilical blood flow.[13] Animals underwent severe variable FHR decelerations with complete UCO for 1 minute duration every 2.5 minutes until the targeted fetal arterial pH of less than 7.0 was detected or 2 hours of severe UCO had been carried out, at which point the repetitive UCOs were terminated. These animals were then allowed to recover for 48 hours following the last UCO. Fetal arterial blood samples were drawn at baseline, at the end of the first UCO of each series (mild, moderate, severe), and at 20 minute intervals (between UCOs) throughout each of the series, as well as at 1, 24, and 48 hours of recovery. For each UCO series blood gas sample and the 24h recovery sample of 0.7 ml of fetal blood was withdrawn, while 4 ml of fetal blood was withdrawn at baseline, at $pH_{nadir}$ < 7.00, and at 1 hour and 48 hours of recovery. The amounts of blood withdrawn were documented for each fetus and replaced with an equivalent volume of maternal blood at the end of day 1 of study.

All blood samples were analyzed for blood gas values, pH, glucose, and lactate with an ABL-725 blood gas analyzer (Radiometer Medical, Copenhagen, Denmark) with temperature corrected to 39.0 °C. Plasma from the 4 ml blood samples was frozen and stored for cytokine analysis, reported elsewhere.

After the 48 hours recovery blood sample, the ewe and the fetus were killed by an overdose of barbiturate (30 mg sodium pentobarbital IV, MTC Pharmaceuticals, Cambridge, Canada). A post mortem was carried out during which fetal sex and weight were determined and the location and function of the umbilical occluder were confirmed. The fetal brain was perfusion-fixed and subsequently dissected and processed for later immunohistochemical study as reported. [14]

Data acquisition and analysis

A computerized data acquisition system was used to record fetal systemic arterial and amniotic pressures and the ECG signal, as described.[7] All signals were monitored continuously throughout the experiment. Arterial and amniotic pressures were measured using Statham pressure transducers (P23 ID; Gould Inc., Oxnard, CA). Arterial blood pressure (ABP) was determined as the difference between instantaneous values of arterial and amniotic pressures. A PowerLab system was used for data acquisition and analysis (Chart 5 For Windows, ADInstruments Pty Ltd, Castile Hill, Australia). Pressures, ECOG and ECG were recorded and digitized at 1000 Hz for further study. For ECG, a 60 Hz notch filter was applied.

R peaks of ECG were used to derive the heart rate variability (HRV) times series. The time series of R-R peak intervals were uniformly resampled at 4 Hz and the RMSSD was then calculated continuously on both original R-R interval time series (with 1 millisecond resolution) and the R-R interval time series resampled at 4 Hz, from each five minutes HRV segment in 2.5 minutes overlapping sliding windows. For the ~6-hour time series, this corresponded to roughly 150 data points.

During UCO series, the point at which hypotensive ABP responses to UCO had been detected by "expert" visual detection were termed ABP "sentinel" values, defined as the time between the onset of such ABP responses to UCO and the time when pH nadir (pH < 7.00) is reached in each fetus.

To detect changes in RMSSD values corresponding to the above sentinel time point in the ABP responses, we used the previously reported algorithm, referred to as Delta point method, based on change point detection.[15] Briefly, Delta point method is a real-time change point detection method, robust to false-alarms, designed to filter a vector of suspected change points. It proceeds by fitting a probabilistic Gaussian process model to the RMSSD time series baseline data and computing online predictions of the RMSSD values within the range of the model. Suspected change-points are declared as significant ($p < 0.05$) deviations from pointwise model predictions and observations. These are viewed as observations of a doubly stochastic Poisson process, with observation rate governed by the Gaussian process model. Based upon this theory, the points are grouped into time intervals, within which the Delta point is selected as the most significant change.

To perform hyper-parameter training, we segmented the data into a 60 point training set, or 2.5-hour training time on the baseline and mild UCO periods of each time series. Our method uses an n = 10 point or 25-minute interval to segment the time series for delta point evaluation. The choice of 10 points or 25-minute interval is to provide a reasonable number of points per interval for the Delta point method, so that a reasonable average may be calculated for the average run in each interval.

We defined a successful detection as the agreement between the Delta point and the sentinel value, with Delta point detection no later than two minutes behind expert detection. In one instance, earlier detection of the trend, approximately 1.5 hours prior to expert visual detection, and subsequent earlier Delta point declaration was considered to be beneficial clinically (Table 1). This demonstrates the effectiveness of the method, suggesting clinical benefits for earlier decision making.

Statistical analysis

The differences in the change point detection at 4 Hz compared to 1000 Hz were evaluated with the Wilcoxon signed-rank test with a P value less than 0.05 considered significant.

**Results**

The physiological characteristics of the experimental groups have been reported.[8,10,11]

Delta point method was able to match the expert prediction with Delta point declaration occurring 14 ± 21 minutes before ABP sentinel time. This corresponded to 92% sensitivity, 86% accuracy and 92% precision.

In the 4 Hz RMSSD time series, the algorithm triggered change point at 45 ± 34 minutes failing to match the expert prediction by yielding 3 times earlier detection times than at 1000 Hz, i.e., producing false positive alarms in 10 out of 14 cases (p = 0.004). This corresponded to 67% sensitivity, 14% accuracy and 18% precision.

The visual inspection of the RMSSD tracings suggested that the overestimation of the baseline FHR variability by RMSSD at the 4 Hz sampling rate is the cause of this false detection phenomenon. To verify this assumption we determined the RMSSD values computed from the 1000 Hz and 4 Hz sampled FHR data sets at baseline and during the UCO series. Confirming our hypothesis, we found a smaller difference in the average normalized RMSSD values during the UCO series compared to the baseline in the 4 Hz data set (0.52 ± 0.16) compared to the 1000 Hz data set (0.85 ± 0.4, p = 0.027).

**Conclusion**

Our findings validate the hypothesis that Delta point method, applied to the FHR-derived HRV measure RMSSD, identifies individual time points of ABP responses to UCO with worsening acidemia by extracting change point information from the physiologically related fluctuations in RMSSD time series. The present findings also show the dependence of this method on high temporal precision of FHR acquisition to capture correctly the physiological fluctuations of FHR at baseline. This is in line with the previous observations in the pregnant sheep model and human fetuses intrapartum.[7,16]

This finding has clinical implications, since high precision HRV can be recorded non-invasively in human fetuses from abdominal ECG.[17,18] Moreover, the present results validate and extend the insight we reported earlier in sheep and human fetuses whereby the reduced sampling rate of FHR acquisition reduces the precision of HRV - derived measures such as RMSSD.[7,16] Here, we show that the Delta point method performs 3-times more precisely in alerting to fetal cardiovascular decompensation when the underlying FHR signal was sampled at the gold standard 1000 Hz rate available with today's fetal ECG monitors rather than at the 4 Hz rate as acquired with the ultrasound monitors.

We had reported consistent changes in fetal brain electrical activity, the electrocorticogram (ECOG), with amplitude suppression and frequency increase during FHR decelerations accompanied by highly correlated pathological decreases in fetal ABP, referred to as adaptive brain shutdown.[3] These changes in ECOG occurred on average 50 minutes prior to attaining a severe degree of acidemia (i.e., fetal arterial pH<7.00). However, we noted a high degree of inter-individual variability in the timing of the onset of these brain electrical and cardiovascular responses. Importantly for the neonatal outcome, we found a relationship between the ensuing neuroinflammation measured by microglial counts, the brain's immune cells, and the timing of the adaptive brain shutdown onset.[14] An individualized and timely detection of the onset of hypotensive responses to worsening acidemia and hence the timing of the adaptive brain shutdown would provide clinically relevant information on the degree of neuroinflammation after birth. Perinatal neuroinflammation has been identified as relevant prognostically not only short-term during early life, but also long-term for adult neurodevelopmental sequelae.[19–26]

We suggest that the robust performance of the algorithm is owed to selecting causally linked phenomena which are reflected in two different time series: RMSSD is known to rise with worsening acidemia due to chemoreceptors activation for example.[6,7] Meanwhile, fetal ABP responses to worsening acidemia deteriorate over time with an initial phase of hypertensive responses during each UCO to compensate for the drop in FHR, followed by the gradual decline of this hypertensive component and eventually hypotensive ABP responses.[3] This is at least partially due to a cardiac decompensation with growing levels of acidemia.[27,28] Acidemia impacts myocardial contractility which decreases cardiac output and ABP. It is plausible to expect that such transition in cardiac behavior will be reflected in HRV, RMSSD in particular, because HRV reflects not only influences of the autonomic nervous system's vagal modulation of the cardiac sinus node activity, it also depends on the intrinsic cardiac rhythm fluctuations and health as evidenced by a decrease in HRV in patients after heart transplants and by presence of HRV as early as in term fetuses of gestational age similar to the present study.[28–32]

The RMSSD time series were highly non-stationary, with four different regimes and three regime changes, corresponding to a baseline period followed by mild, moderate and severe UCO series. Each time series was characterized by seemingly randomly occurring (in terms of timing of the individual onset) increase in RMSSD values at different time points during the moderate UCO series and at the start of the severe UCO series. This event manifested as an increasing trend in RMSSD values, which counter-intuitively emerged as a period of relative stationarity for the time series. The Delta point algorithm effectively declared these points as the change point of clinical importance. Overall, we found the Delta point algorithm's predictions to be reliable even in the instances when the signals were noisy.[15] This is based on the tests of the algorithm in various data sets as published [15] and on our observation that here, to mimic the online recording situation, no correction for ectopies or non-sinus rhythms was undertaken on RMSSD as is usually done for HRV offline processing.[31] To our knowledge, no comparable statistical methods for FHR analysis exist.

In conclusion, the novelty of the current work is that our algorithm permits statistical-level predictions about concomitant changes in individual FHR tracings which alert about fetal cardiovascular decompensation, an important mechanistic prequel to brain injury.

**Acknowledgments**
The authors gratefully acknowledge Dr. Bryan Richardson and his Perinatal Research at the University of Western Ontario for the original design of the animal experiments that enabled the acquisition of the dataset underlying the present study. MGF is funded by the Canadian Institutes for Health Research (CIHR).

**Tables**

**Table 1.**

| Group | Animal | Sentinel | 1000 Hz detection | 4Hz detection | 1000 Hz delta | 4 Hz delta |
|---|---:|---|---|---|---|---|
| H_UCO | 8003 | 15:56:00 | 15:49:00 | 15:52:00 | 0:07 | 0:04 |
| H_UCO | 473351 | 13:38:00 | 13:28:00 | 13:04:00 | 0:10 | 0:34 |
| H_UCO | 473362 | 11:05:00 | 11:03:00 | 11:35:00 | 0:02 | <span style="color:red">0:30</span> |
| H_UCO | 473376 | 12:36:00 | 12:38:00 | 11:59:00 | <span style="color:red">0:02</span> | 0:37 |
| H_UCO | 473726 | 12:04:00 | 11:50:00 | 11:54:00 | 0:14 | 0:10 |
| N_UCO | 461060 | 12:42:00 | 12:31:00 | 12:21:00 | 0:11 | 0:21 |
| N_UCO | 473361 | 12:51:00 | 12:36:00 | 12:16:00 | 0:15 | 0:35 |
| N_UCO | 473352 | 13:17:00 | 12:53:00 | 12:06:00 | 0:24 | 1:11 |
| N_UCO | 473377 | 12:12:00 | 12:14:00 | 12:50:00 | <span style="color:red">0:02</span> | <span style="color:red">0:38</span> |
| N_UCO | 473378 | 13:22:00 | 13:09:00 | 12:09:00 | 0:13 | 1:13 |
| N_UCO | 473727 | 11:03:00 | 11:10:00 | 11:08:00 | <span style="color:red">0:07</span> | <span style="color:red">0:05</span> |
| N_UCO | 5054 | 12:53:00 | 11:27:00 | 11:19:00 | 1:26 | 1:34 |
| N_UCO | 5060 | 11:26:00 | 11:24:00 | 10:29:00 | 0:02 | 0:57 |
| N_UCO | 473360 | 13:59:00 | 13:52:00 | 11:55:00 | 0:07 | 2:04 |

*Sentinel*, time of detecting the onset of pathological ABP decreases during UCOs by an expert (visual analysis);
*1000 Hz and 4 Hz detection*, times of detecting the same using the change point algorithm on RMSSD data derived from 1000 Hz or 4 Hz sampled ECG;
*1000 Hz delta and 4 Hz delta*, time difference (sentinel - 1000 Hz or sentinel - 4 Hz) between expert and change point algorithm detection performance: detection by the algorithm preceded in most cases the expert detection, 14 ± 21 minutes and 45 ± 34 minutes, respectively;
*Red font*, cases when the algorithm detection happened *after* the expert detection; note that in the case of 4 Hz delta, 2 out of 3 instances the detection was more than 30 min too late compared to ~3 min too late in the three cases at 4 Hz.

**Figures**

**Fig. 1**: Representative example from the experimental data (animal ID 473378). The RMSSD time series derived from 1000 Hz (blue) and 4 Hz (orange) sampled ECG are displayed superimposed in the top panel, with declared change points from the BOCPD algorithm and the Sentinel (expert detection) marked with arrows. Sequential fetal arterial pH measurements are indicated. Experimental stages are demarcated by background colors; short black bars over X axis indicate the zoomed-in segments shown in the bottom panel.
Bottom panels show fetal heart rate (FHR, bpm), fetal arterial blood pressure (ABP, mmHg) and umbilical cord pressure (UCP, mmHg) indicating when UCO were triggered (increasing UCP). Note the failure of the change point algorithm to detect the sentinel time point correctly (i.e., around the Sentinel time point) when using 4 Hz - derived RMSSD signal: the detection occurs 1 hour earlier than with the 1000 Hz signal. This is due to unphysiological fluctuations in FHR variability at baseline as we demonstrated in [7,16].

**Figure 1.**

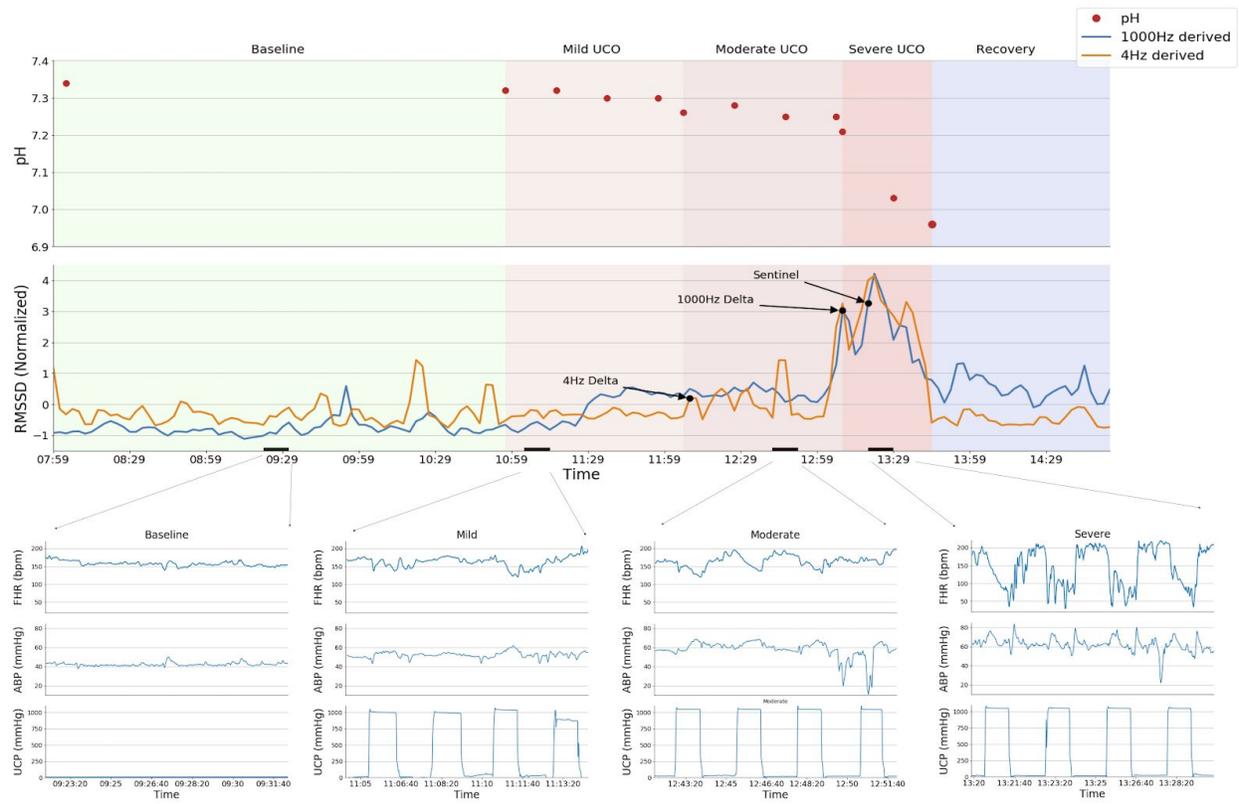